\documentclass[aps,final]{revtex4}
\usepackage{amssymb,amsmath}
\usepackage{graphicx}

\newcommand{\ave}[1]{\left \langle {#1} \right \rangle}

\DeclareMathOperator{\var}{var}

\begin{document}

\title{A simple model for Carnot heat engines}

\author{Jacques Arnaud}
\affiliation{Mas Liron, F30440 Saint Martial, France},
\author{Laurent Chusseau}
\affiliation{Institut d'\'Electronique du Sud, UMR n$^\circ$5214 au CNRS, Universit\'e Montpellier II, F34095 Montpellier, France},
\author{Fabrice Philippe}
\altaffiliation[Also at ]{Universit\'e Montpellier 3, route de Mende, F34199 Montpellier, France}
\affiliation{LIRMM, UMR n$^\circ$5506 au CNRS, 161 rue Ada, F34392 Montpellier, France}

\date{\today}

\begin{abstract}

We present a (random) mechanical model consisting of two lottery-like reservoirs at altitude $E_h$ and $E_l<E_h$, respectively, in the earth's gravitational field. Both reservoirs consist of $N$ possible ball locations. The upper reservoir contains initially $n_h\le N$ weight-1 balls and the lower reservoir contains initially $n_l\le N$ weight-1 balls. Empty locations are treated as weight-0 balls. These reservoirs are being shaken up so that all possible ball configurations are equally likely to occur. A cycle consists of exchanging a ball randomly picked from the higher reservoir and a ball randomly picked from the lower reservoir. It is straightforward to show that the efficiency, defined as the ratio of the average work produced to the average energy lost by the higher reservoir is $\eta=1-E_l/E_h$. We then relate this system to a heat engine. This thermal interpretation is applicable only when the number of balls is large. We define the entropy as the logarithm of the number of ball configurations in a reservoir, namely  $S(n)=\ln[N!/n!(N-n)!]$, with subscripts $h,l$ appended to $S$ and to $n$. When $n_l$ does not differ much from $n_h$, the system efficiency quoted above is found to coincide with the maximum efficiency $\eta=1-T_l/T_h$, where the $T$ are absolute temperatures defined from the above expression of $S$. Fluctuations are evaluated in Appendix A, and the history of the Carnot discovery (1824) is recalled in Appendix B. Only elementary physical and mathematical concepts are employed.

\end{abstract}

\maketitle

\section{Introduction}

The purpose of this paper is to give newcomers to the field of Thermodynamics a feel for the concepts involved in a simple manner. The paper is self-contained and employs only elementary mathematics. The results are derived for a particular urn (or bag, or reservoir) model, related to the one introduced in 1907 by P.~Ehrenfest.\cite{Ehrenfest1907, Guemez1989, ISI:000253989000010} Our model consists of two reservoirs with $N$ possible ball locations each, containing respectively $n_l$ and $n_h$ weight-1 balls. This model is directly applicable to Otto heat engines (in which the working agent parameter stays fixed when in contact with either bath) employing as a working agent two-level atoms. 
(For an exhaustive discussion concerning quantum heat engines see Quan.\cite{Quan2009}) Obviously, classical heat engines may of course be considered as special cases of quantum heat engines. 

In section \ref{exchange} we describe in more detail our model, and evaluate the average work produced, the average energy lost by the upper reservoir, and the efficiency. The general properties of heat engines are recalled in Section \ref{heat}. The relationship between our urn model and the properties of heat engines is discussed in Section \ref{entropy}. The reservoirs absolute temperatures $T_l,T_h$, respectively, are defined in the limit of large ball numbers. When $n_l\approx n_h$ the system efficiency and the average work obtained in Section \ref{exchange} tend to coincide with the expression for the efficiency and average work given by Carnot for an ideal heat engine. The \emph{fluctuations} of the work produced are evaluated in Appendix A. A discussion of the history of the Carnot discovery is in Appendix B. 

It is well-known that heat exchange between two contacted bodies may be described by an urn model (see Section \ref{heat}). The present model employing two urns at different altitudes, and the comparison that we make with heat engines, has not, to our knowledge, been presented before.

It seems to us that mathematically-minded students would better understand the entropy concept from the present model than from classical Thermodynamics, because empirical results (such as the irreversibility of thermal contacts) are not needed in our discussion. For other students, temperature is a primary intuitive concept, and they may prefer an introduction based on gas-filled cylinders of variable length instead.

\section{Exchange of balls between two reservoirs}\label{exchange}

\begin{figure}
\centering
\includegraphics[scale=1]{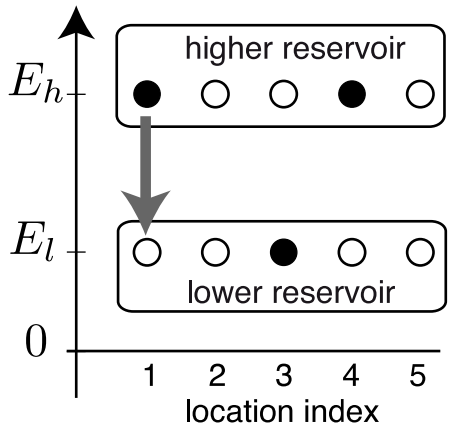}
\caption{Schematic representation of an engine that converts potential energy into work. The figure represents two lottery-like reservoirs located at altitudes, $E_l$ and $E_h\ge E_l$, respectively, with $N$ possible ball locations labeled 1,2,3...$N$ ($N=5$ in the figure). The number of weight-1 balls (black circles) is $n_l$ in the lower reservoir, and $n_h$ in the higher reservoir (with $n_l=1,~n_h=2$ in the figure). Open circles may be viewed as weight-zero balls. For each reservoir, every ball configuration is equally likely to occur considering that the energies are the same. The complete figure should therefore consist of $5\times 10 = 50$ similar figures exhibiting all the possible system configurations. Balls may be transferred from one reservoir to the other in location 1 only. If there is a ball in the upper reservoir at that location and none in the lower one (as is the case in the figure), the ball gets transferred from the upper reservoir to the lower one, thereby delivering energy. Conversely, if there is a ball in the lower reservoir and none in the upper one the ball gets transferred from the lower reservoir to the upper one, thereby absorbing energy. In some limits, the efficiency is the same as for a Carnot cycle. }
\label{fig}
\end{figure}

We consider a system consisting of two reservoirs at altitudes $E_{l}$ and $E_{h}>E_{l}$ respectively, with respect to some lower reference level, in the earth's gravitational field. There are $N$ possible ball locations in each reservoir that we label: 1,2...$N$, as shown in Fig. \ref{fig}. A single ball is allowed in any given location. The lower reservoir contains $n_l\le N$ weight-1 balls and the higher reservoir contains $n_h\le N$ weight-1 balls. By ``weight-1 ball,'' we mean an object with a weight of 1 N, that is, with a mass of approximately 0.1 kg. In order to lift such an object by 1 m in the earth's gravitational field an energy of 1 J is required. Conversely, such an object delivers an energy of 1 J whenever its altitude gets lowered by 1 m. These energies may be removed or added to some external device with the help, for example, of cords and pulleys, or electrical motors (or generators). We leave the nature of these mechanisms unspecified in the discussion that follows. Note, however, that a particularly interesting mechanism is the action of resonant optical fields on two-level atoms. They may convert upper-state atoms into lower-state atoms and conversely, thereby receiving or delivering energy.

In Figure~\ref{fig} we have $N=5$, $n_l=1$ and $n_h=2$. In the higher reservoir there are 10 distinguishable ball configurations and in the lower reservoir there are 5 distinguishable configurations (the general formula for the number of configurations will be given later on). Considering the two reservoirs together there are therefore $5\times 10$=50 distinguishable configurations. These 50 configurations correspond to the same energy. If one assumes that the reservoirs are being shaken up frequently (at least more frequently than the occurrence of a ball-exchange event), all possible ball configurations are equally likely to occur. 

We consider ball displacements from the lower and higher reservoirs at a single location, say the left-most location, labeled 1. One may draw all the possible configurations and evaluate by inspection quantities of interest such as the average work produced by the system. For example, if $N=3,~n_l=1,~n_h=2$, the nine distinguishable configurations are 
\begin{align}\nonumber
\begin{array}{c}\circ\bullet\bullet\\\bullet\circ\circ\\-1\end{array}\;
\begin{array}{c}\bullet\circ\bullet\\\bullet\circ\circ\\0\end{array}\;
\begin{array}{c}\bullet\bullet\circ\\\bullet\circ\circ\\0\end{array}\;
\begin{array}{c}\circ\bullet\bullet\\\circ\bullet\circ\\0\end{array}\;
\begin{array}{c}\bullet\circ\bullet\\\circ\bullet\circ\\1\end{array}\;
\begin{array}{c}\bullet\bullet\circ\\\circ\bullet\circ\\1\end{array}\;
\begin{array}{c}\circ\bullet\bullet\\\circ\circ\bullet\\0\end{array}\;
\begin{array}{c}\bullet\circ\bullet\\\circ\circ\bullet\\1\end{array}\;
\begin{array}{c}\bullet\bullet\circ\\\circ\circ\bullet\\1\end{array}
\end{align}
The number below each configuration is the energy produced over a cycle, setting for brevity $E\equiv E_h-E_l=1$. Indeed, restricting our attention to the first location, labeled 1, and, for example, the first configuration shown above, we note that there is an empty location in the higher reservoir and a ball in the lower reservoir. In that case, a cycle consists of transferring the lower-reservoir ball into the higher empty location. The system then delivers an energy equal to $-1$, that is, absorbs an energy equal to 1. From the above numbers, we calculate that the average energy produced, called ``average work,'' is $\ave{W}=(-1+0+0+0+1+1+0+1+1)/9=1/3$. This pedestrian method also provides the variance of the work produced.

We now turn to an equivalent but more convenient picture (see the abstract), involving the probability of picking up a ball from a reservoir. It is then convenient to suppose that empty locations are occupied by weight-0 balls. Such balls, of course, do not carry any energy when displaced. A cycle consists of exchanging two balls (of weight 0 or 1), one randomly picked from the lower reservoir, and one randomly picked from the higher reservoir. The probability of picking a weight-1 ball from the  lower reservoir is $l\equiv n_l/N$ and the probability of picking up a weight-1 ball from the  higher reservoir is $h\equiv n_h/N$. As we discuss below, the average energy produced is $\ave{W}=h-l$. In the above example (namely $N=3,~n_l=1,~n_h=2$), $l=1/3, ~h=2/3$, we obtain $\ave{W}=2/3-1/3=1/3$, which coincides with the previous result. The equivalence between the two methods is general.

The total (potential) energy of  a reservoir at altitude $E$ (with respect to some arbitrarily selected level) containing $n$ weight-1 balls is obviously $Q=nE$ (kinetic energy being not considered, the total energy coincides with the potential energy). The letter $Q$ is employed anticipating a correspondence with heat. When a weight-1 ball is added to a reservoir at altitude $E$ the reservoir energy is incremented by $E$. On the other hand, if a ball is randomly picked up from the $N$ locations of a reservoir containing $n$ weight-1 balls, the probability that this ball has weight 1 is clearly $n/N$. Accordingly,  if the picked up ball is subsequently carried to a reservoir at altitude $E$, the latter reservoir average energy is incremented by $\Delta Q=E~ n/N$. The word ``average'' will henceforth be omitted in the main text since only average values are considered.

Consider now two such reservoirs. One at altitude $E_l$ (lower reservoir) and containing $n_l$ weight-1 balls. The other at altitude $E_h$ (higher reservoir) containing $n_h$ weight-1 balls. A cycle consists of exchanging two randomly-picked balls between the two reservoirs. From what has just been said and setting $l\equiv n_l/N$, $h\equiv n_h/N$, the energies added to the lower and higher reservoirs read respectively 
\begin{align}
\label{a} 
\Delta Q_{l}=E_{l}(h-l), \qquad 
\Delta Q_{h}=-E_{h}(h-l).
\end{align}
The work performed follows from the law of conservation of energy
\begin{align}\label{b}
 W=-\Delta Q_{l}-\Delta Q_{h}=(E_{h}-E_{l})(h-l).
\end{align}
The engine efficiency, defined as the ratio of the work performed $W$ and the energy $-\Delta Q_{h}$ lost by the higher reservoir, is therefore 
\begin{align}
\label{c} 
\eta\equiv \frac{W}{-\Delta Q_{h}}=1-\frac{E_{l}}{E_{h}}.
\end{align}

In the next section we recall well-known properties of heat engines. In the subsequent section, we show that when $h\approx l$ the efficiency given in Eq.~\eqref{c} coincides with the Carnot efficiency and the work given in Eq.~\eqref{b} coincides with the expression given by Carnot. 

Note that after a cycle the number of weight-1 ball in a reservoir may be incremented by -1, 0 or 1. The next cycle operates therefore with different values of the $l,~h$ parameters. We will not consider the evolution of the work produced cycle after cycle. Indeed we are interested in a comparison with heat engines operating between two baths whose heat capacity is so large that their temperatures do not vary significantly. Likewise, in the present reservoir model, one may suppose that the number of balls in each reservoir is so large that their change after any number of cycles is insignificant.

\section{Heat engines}\label{heat}

The purpose of the present section is to recall some well-known facts about heat and heat engines  (for a detailed discussion see for example Ref.~\onlinecite{PRE9}).  

It is empirically known that when two bodies at different temperatures are contacted, they eventually reach the same temperature. This observation can be verified using only the intuitive concept of temperature. Classically, this fact may be interpreted by supposing that some heat (energy) is flowing from the high-temperature body into the low-temperature body, but that the converse never occurs. A similar observation can be made with two urns (see the figure with $E_l=E_h$). Suppose that the urns contain $N$ balls each that have either weight-1 (black) or weight-0 (white). If randomly picked balls are repeatedly exchanged between the two urns, eventually the ratios of black and white balls become nearly the same in both urns, irrespectively of the initial conditions. For the analogy with the phenomenon of thermal contact to hold, one must define the temperature of an urn as a monotonically increasing function of the ratio of black to white balls [for an exact expression, see Eq.~\eqref{f} with $E>0$ and $n=n_{black}, ~N-n=n_ {white}$].

It is also an empirical fact that when the energy originating, for example, from a dropping weight is dissipated into a body the body heats up. According to the law of conservation of energy, one thus presumes that a hot body contains an energy called  ``heat.'' The fact that this heat cannot be converted back into usable energy is one form of the second-law of Thermodynamics. In order to get usable energy from heat one needs two bodies at different temperatures, one of them being called the high-temperature bath and the other the low-temperature bath. If the low-temperature bath is at zero absolute temperature, all the heat may in principle be converted back into usable energy, but this is not the case in general. Heat is usually pictured as the average kinetic energy of randomly moving atoms, according to Bernoulli and Maxwell pictures. For example, the average kinetic energy of an helium atom in a gas is $3T/2$ in appropriate units, where $T$ denotes the gas absolute temperature. We will not discuss further this conventional model because in the present paper we instead represent heat by a random \emph{potential} energy. Consider for example the first location of the higher reservoir in Fig.\;\ref{fig}. For the configuration shown in that figure the energy is equal to $E_h$ since a weight-1 ball at altitude $E_h$ is present. But for other configurations (not shown in the figure), the energy may be zero. On the average, the higher reservoir energy is easily found to be (2/5)$E_h$.

Let us now recall how, concretely, usable energy may be retrieved from two baths at different temperatures. What one needs is a ``working agent,'' which may be any piece of material whose properties can be changed by varying a parameter. An example is an helium-filled cylinder with a piston, of length $\epsilon$, where $\epsilon$ is the parameter. Another example (more relevant to this paper discussion) is a collection of two-level atoms. We suppose that the separation in energy $\epsilon$ of the two levels may be varied through, e.g., the application of an electrical field. It is the external agent that causes the parameter $\epsilon$ to vary that collects, or delivers, the energy. A typical closed cycle consists of putting the working agent with parameter $\epsilon_1$ in contact with the low-temperature bath, and slowly varying this parameter to $\epsilon_2$. The working agent is then carried from the low-temperature bath to the high-temperature bath while the parameter is slowly changed to $\epsilon_3$. The parameter is then changed to $\epsilon_4$. The working agent is finally carried back to the low temperature bath while the parameter recovers its initial $\epsilon_1$-value. A closed cycle is thus defined by the nature of the working agent and four parameter values. If the parameter $\epsilon$ does not vary at all, obviously no energy is delivered or received. The cycle then simply transfers heat from the high-temperature bath to the low-temperature bath, and is analogous to a thermal contact. 

A more interesting situation is the Otto cycle that describes an idealized form of the gasoline engine (discovered by Beau de Rochas in 1862 and Otto in 1876). In that case $\epsilon_2=\epsilon_1$ and $\epsilon_4=\epsilon_3$, meaning that the parameter does not vary when the working agent is in contact with either bath. The parameter varies only during the adiabatic transitions from one bath to the other. This cycle may deliver work or receive work (heat pump) for appropriate choices of the parameters. Usually it does not achieve the maximum (Carnot) efficiency. It does so approximately, however, when $\epsilon_4\approx \epsilon_2$. For the exact description of Carnot heat engines, see the generalization of the ball model in Ref.~\onlinecite{ourPRE}.

In the celebrated Carnot cycle the parameters are so chosen that the temperature of the working agent is nearly the same as the bath temperature when contacted with it. Then the efficiency is $\eta=1-T_l/T_h$, and the work produced is $W=(T_h-T_l)S$, where $S$ denotes the entropy transferred from the high-temperature bath to the low-temperature bath. These concepts will be made clearer in the next section.

To summarize, whenever we have at our disposal two baths at different temperatures, one may always find a heat engine that delivers energy. But many kind of cycles fail to deliver energy even though $T_h>T_l$. As a matter of fact they may instead \emph{absorb} energy and act as heat pumps. As we shall see, our potential-energy model is fully consistent with the above well-known considerations.

\section{Entropy and temperature}\label{entropy}

We consider again reservoirs containing $N$ locations and $n$ weight-1 balls and $N-n$ weight-0 balls. To relate this device to \emph{heat} engines, let us first recall that the number of ball configurations in a reservoir is $N!/n!(N-n)!$. For example, if $N=3$ and $n=1$, there are 3!/(1!2!)=3 configurations, namely
$(\bullet\circ\circ)$, $(\circ\bullet\circ)$ and $(\circ\circ\bullet)$. Next, we define the entropy as the logarithm of the number of configurations, the Boltzmann constant being set equal to unity, that is, for a reservoir, 
\begin{align}
\label{d} 
S(n)=\ln \left(\frac{N!}{n!(N-n)!}\right).
\end{align}
Note that 
\begin{align}
S(n+1)-S(n)&=\ln \left(\frac{N!}{(n+1)!(N-n-1)!}\right)-\ln \left(\frac{N!}{n!(N-n)!} \right)\nonumber \\
&=\ln \left(\frac{N-n}{n+1} \right)\approx \ln\left(\frac{N}{n}-1\right),\label{e}
\end{align}
for large $n$.

The absolute temperature of a reservoir is then defined as 
\begin{align}
\label{f} 
T(n)&=\frac{Q(n+1)-Q(n)}{S(n+1)-S(n)}\approx \frac{E}{\ln(\frac{N}{n}-1)}.
\end{align}
Temperature is an \emph{intensive} quantity. For example, the temperature of two identical bodies at temperature $T$, considered together, is again $T$. Because heat has the nature of an energy and is an extensive quantity, it is required that $S$ be also an extensive quantity. Since the number of configurations in two separate bodies is the \emph{product} of the configurations (for each configuration of one body one must consider all the configurations of the other body) and the logarithmic function has the property that $\ln(ab)=\ln(a)+\ln(b)$, the above definitions do ensure that $T$ be an intensive quantity. Note that we have chosen a temperature unit such that the Boltzmann constant $k_B$ be unity. By doing so the distinction between extensive and intensive quantities drops out of sight. For example, the energy of a single-mode oscillator $E=k_B T$ reads in our notation $E=T$. However, the distinction may be restored, while keeping $k_B=1$, by writing $E=T\times$ the number of modes. The number of modes depends on volume, while $T$ does not. Note that the temperature is positive if $n<N/2$.

The cycle efficiency given in Eq.~\eqref{c} may now be written in terms of temperatures as
\begin{align}
\eta=1-\frac{E_{l}}{E_{h}}
=1-\frac{T_l}{T_h}\frac{\ln(\frac{1}{l}-1)}{\ln(\frac{1}{h}-1)}\label{g} .
\end{align}
Thus, when $l\approx h$, the last fraction in the above equation drops out and the Carnot efficiency is indeed obtained.  In the limit $l\approx h$ the work $W$ produced per cycle is very small. However, one may always add up the work contributions of any number of similar devices having the same reservoir temperatures (but possibly different values of $E, ~ n$), and achieve any specified work at the Carnot efficiency.

The ball exchange discussed above may increment the reservoir entropies. The number of weight-1 balls in a reservoir may indeed be incremented by one, remain the same, or be decremented by one. From what was said before, the probability that a weight-1 ball be transferred from the high reservoir to the lower one is $h\equiv n_h/N$, and the probability that a weight-1 ball be transferred from the low reservoir to the higher one is $l\equiv n_l/N$. Since these events are independent, the lower reservoir entropy increment reads
\begin{align}\label{h}
\Delta S_l=h(1-l)[S(n_l+1)-S(n_l)]+l(1-h)[S(n_l-1)-S(n_l)].
\end{align}
Using Eq.~\eqref{e} we obtain
\begin{align}\label{i}
\Delta S_l=(h-l)\ln\left(\frac{1}{l}-1\right) .
\end{align}
The increment of the higher reservoir entropy is obtained by exchanging the $h$ and $l$ labels in the above expression, that is
\begin{align}\label{j}
\Delta S_h=-(h-l)\ln\left(\frac{1}{h}-1\right) .
\end{align}

We thus find that, in the limit $n_l/N\approx n_h/N$ (or $l\approx h$), 
$\Delta S_l\approx -\Delta S_h$ so that there is no net entropy produced. 
Entropy is just carried from the higher reservoir to the lower one.
The Carnot expression for the work recalled in Section \ref{heat} may thus be written as
\begin{align}
W=(T_h-T_l)\Delta S_l&\approx \left( \frac{E_h}{\ln\left(\frac{1}{h}-1\right)}- \frac{E_l}{\ln\left(\frac{1}{l}-1\right)} \right)\Delta S_l  \nonumber \\
&\approx (E_h-E_l)(h-l) \label{k},
\end{align}
so that the Carnot general formula for $W$ indeed coincides with the expression for the work performed per cycle evaluated for our model from simple reasoning. 
More precisely, the ratio of the total entropy produced $\Delta S_l+\Delta S_h$ to the work produced $W$ tends to zero as $h\to l$, see Eq.~\eqref{mbis}.

The final expression tells us that the engine, according to our model, delivers work only if the terms $E_h-E_l$ and $h-l$ are both positive or both negative. But since $E_h>E_l$ by convention, this implies that we must have $h>l$. Going back to the expression of the temperature in Eq.~\eqref{f}, and remembering that the $\ln(.)$ function is a monotonically increasing function of its argument, we find that work may be produced only if $T_h>T_l$. Whenever $T_h>T_l$ there exist heat engines that may deliver work. But this is not so for every heat engine. In particular, one may consider the properties of an Otto heat engine that stops delivering work (and turns into a heat pump) when a condition similar to our $l=h$ condition holds, even though $T_h>T_l$.

Conventional heat engines operate with two large baths, or reservoirs, one hot and one cold. Because these baths are not infinite in size, cycle after cycle, the hot bath cools down and the cold bath warms up. Eventually no work is being produced. The same situation occurs in our model. After a very large number of cycles the values of $h$ and the value of $l$ tend to coincide and no work is being produced any more. Because the reservoir temperatures do not equalize, however, one may say that the system has then reached a state of equilibrium, but not a state of \emph{thermal} equilibrium. This is not a peculiarity of our model, but a general property of some heat engines. 

\section {Conclusion}
   
We have seen that heat engines may be equivalent to random mechanical engines of a special kind. Precisely, the model consists of two reservoirs having $N$ locations, with $n_l,~n_h$ weight-1 balls, at different altitudes. The only concepts involved in the present paper are those of potential energy and of uniform probability. We have shown that the efficiency and work in our model coincide with the Carnot expressions in the limit where $n_l\approx n_h$. Full Carnot cycles may be generated out of this elementary configuration.

\appendix

\section{Fluctuations}

Some readers may not be particularly interested in the fluctuations of the quantities of major interest considered in the main text, namely the work produced and the high-temperature reservoir heat loss. However, fluctuations are important in some applications. We show in this appendix that the variance of these quantities can be readily obtained from the ball model. 

Recall that in our model a cycle consists of exchanging simultaneously a ball from the higher reservoir (at altitude $E_h$ and containing $n_h$ weight-1 balls and $N-n_h$ weight-0 balls) and a ball from the lower reservoir (at altitude $E_l$ and containing $n_l$ weight-1 balls and $N-n_l$ weight-0 balls). The probability that a weight-1 ball be picked up from the higher reservoir is $h\equiv n_h/N$. The probability that a weight-1 ball be picked up from the lower reservoir is $l\equiv n_l/N$. The two events are independent. 

Setting $E\equiv E_h-E_l$, we have seen in the main text that the average work produced per cycle is $\ave{W}=E(h-l)$. We now evaluate $\ave{W^2}$. The probability that a weight-1 ball falls and none is raised is $h(1-l)$. If this event occurs, the work performed squared is equal to $E^2$. Conversely, the probability that a weight-1 ball is raised and none falls is $l(1-h)$. If this event occurs, the work performed squared is again equal to $E^2$. Because the two other cases produce no work, it follows that $\ave{W^2}=E^2[h(1-l)+l(1-h)] $. Therefore, the variance of the work produced reads
\begin{align}\label{l}
\var(W)&\equiv \ave{W^2}-\ave{W}^2\nonumber\\
&=E^2[h(1-l)+l(1-h)-(h-l)^2]=E^2[h(1-h)+l(1-l)] .
\end{align}
In the limit $h\approx l$ considered in the main text, we have 
\begin{align}\label{lbis}
\var(W)\approx2E^2 l(1-l).
\end{align}

Let us now consider the total entropy produced $\Delta S\equiv \Delta S_l+\Delta S_h$. When a weight-1 ball is being transferred from the high reservoir to the lower one and none from the low reservoir to the higher one, an event that occurs with probability $h(1-l)$, the increment of $S_l$ is, according to Eq.~\eqref{e}, $\Delta S(n_l+1)-\Delta S(n_l)=\ln(\frac{1}{l}-1)$, and the increment of $S_h$ is $\Delta S(n_h-1)-\Delta S(n_h)=-\ln(\frac{1}{h}-1)$. It follows that the increment in total entropy is $\ln(\frac{\frac{1}{l}-1}{\frac{1}{h}-1})$ with probability $h(1-l)$. When a ball is being transferred from the low reservoir to the higher one and none from the high reservoir to the lower one, an event that occurs with probability $l(1-h)$, the increment of $S_l$ is, according to \eqref{e}, $\Delta S(n_l-1)-\Delta S(n_l)=-\ln(\frac{1}{l}-1)$, and the increment of $S_h$ is $\Delta S(n_h+1)-\Delta S(n_h)=\ln(\frac{1}{h}-1)$. It follows that the increment in total entropy is $\ln(\frac{\frac{1}{h}-1}{\frac{1}{l}-1})$ with probability $l(1-h)$.

The average increment in total entropy is therefore
\begin{align}
\ave{\Delta S}&=h(1-l)\ln \left( \frac{\frac{1}{l}-1}{\frac{1}{h}-1} \right) +l(1-h)\ln \left(\frac{\frac{1}{h}-1}{\frac{1}{l}-1} \right)\nonumber\\
&=(h-l) \ln \left(\frac{\frac{1}{l}-1}{\frac{1}{h}-1}\right) \geqslant 0.\label{m}
\end{align}
As was said in the main text, when $h\approx l$, $\Delta S\approx 0$ and the system tends to be reversible and to achieve the highest efficiency. Note that the entropy increment is non-negative for both a heat engine ($h>l$) and a heat pump ($l>h$). More precisely, noting that to first order in $\delta \equiv h-l$ we have $\ln[(1/l-1)/(1/h-1)]\approx \delta/[l(1-l)]$, and thus
\begin{align}
\ave{\Delta S}\approx\frac{\delta^2}{l(1-l)}.\label{mbis}
\end{align}
The whole model presented makes sense because the generated entropy is proportional to $\delta^2$ while the work produced is proportional to $\delta$, so that, for small $\delta$, near reversibility does not imply vanishing work.

Finally, we evaluate the variance of the total entropy increment. From the above expressions, it follows that 
\begin{align}
\ave{(\Delta S)^2}&=h(1-l)\left[\ln \left(\frac{\frac{1}{l}-1}{\frac{1}{h}-1}\right)\right]^2+l(1-h)\left[\ln \left(\frac{\frac{1}{h}-1}{\frac{1}{l}-1}\right)\right]^2\nonumber\\
&=(h+l-2lh)\left[ \ln \left(\frac{\frac{1}{l}-1}{\frac{1}{h}-1}\right) \right]^2,\label{n}
\end{align}
and the variance reads
\begin{align}
\var(\Delta S)&\equiv \ave{(\Delta S)^2}-\ave{\Delta S}^2\nonumber\\
&=(h+l-2lh-(h-l)^2)\left[ \ln \left(\frac{\frac{1}{l}-1}{\frac{1}{h}-1}\right)\right]^2\nonumber\\
&=\left[(h(1-h)+l(1-l)\right]\left[ \ln\left(\frac{\frac{1}{l}-1}{\frac{1}{h}-1}\right)\right]^2,\label{o}
\end{align}
which vanishes, as well as the average entropy produced, when $h\approx l$. To first order in $\delta\equiv h-l$, we have
\begin{align}
\var(\Delta S)\approx 2\ave{\Delta S},\label{p}
\end{align}
a remarkably simple result.
This result has been presented before in Ref.~\onlinecite{ejp02} just after Eq.~(18), and also in previous works. This agreement shows that the properties of our model are, at least up to a point, generic, that is, generally applicable to heat engines.

\section{Brief history of the Carnot discovery}\label{history}

The motivation for introducing the present account of Carnot discoveries is that, in spite of the efforts of a number of motivated scientists (see below), they remain insufficiently appreciated. 

The Carnot theory, which appeared in a book in 1824 and in unpublished notes, established both the first and the second laws of Thermodynamics. This fact has been pointed out by a number of authors who took the trouble of looking carefully at what Carnot actually said, clarifying  the terminology employed, up-dating the system of units, and correcting minor errors in the experimental data. One of these authors is the Nobel-prize winner A. Kastler.\cite{Kastler1976} We translate from Kastler paper: ``Had Sadi Carnot lived longer [...] he would be considered today not only as the author of the Carnot principle (called by Clausius second principle of Thermodynamics) but also as the author of the first principle of that science.'' Another author is the russian scientist V.M. Brodiansky.\cite{PRE10} At the end of his book, p.~228 and 229, the author lists ten major achievements of Carnot. The first reads ``Carnot is the first to formulate the second principle of thermodynamics'' and the eighth says ``He was among the firsts to formulate strictly the law of equivalence between  heat and work, and the first to calculate with sufficient accuracy its numerical value.'' 

One reason for the current misunderstanding is that part of the Carnot contribution appeared in print only decades after his early death. A second one is that his work was popularized by Clapeyron in a partly erroneous manner. A third one is the unfortunate use by Carnot of the word ``calorique'' to designate what Clausius later on called ``entropy.'' The word ``calorique'' had been formerly employed by Lavoisier to designate some hypothetical heat substance. Clausius and Lord Kelvin, though highly appreciative of the Carnot work, missed part of his contribution because the notes mentioned above were not available to them. Let us cite also the paper by La~Mer who expressed himself forth-fully as follows: ``Unless the view-point that the Carnot theory is accurate is adopted, one is placed in the position of maintaining that Carnot succeeded in demonstrating some of the most fundamental and profound principles of physical science by the most masterly display of scientific double-talk that has ever been perpetrated upon the scientific world. This view is untenable.\cite{Mer1954}'' Much clarification is due to Hoyer.\cite{hoyer} The historian of science R.~Fox says: ``Until recently there were very few studies concerning [the physics of Carnot reflexions]. Thanks to the work of Hoyer, we now have papers on the logical implications of the Carnot theory, and its analogy with modern thermodynamics [...]. It is not at all obvious to understand how Carnot [discovered the mechanical equivalent of heat]. Hoyer examined this question in two important papers. His articles provide complete references to earlier attempts [...]. He explains the exactness of Carnot calculation (which is even more striking if one uses modern values for the specific heats) by noticing that the Carnot theory is entirely accurate.''

As far as the first law of thermodynamics is concerned, let us quote Carnot:\cite{EJP3} ``Heat is nothing but motive power, or rather another form of motion.  Wherever motive power is destroyed, heat is generated in precise proportion to the quantity of motive power destroyed; conversely, wherever heat is destroyed, motive power is generated.'' Carnot calculated that 1 calorie of heat is equivalent to 3.27~J, instead of the modern value: 4.18~J.  

As far as the second law is concerned, the key fact is that engine efficiencies reach their maximum value when they are reversible. Carnot reached this conclusion from the consideration that energy cannot be obtained for free. He therefore looked for heat processes that could work in a reversed manner, ending up with the celebrated ``Carnot cycle.'' Slow processes are reversible, with the exception of thermal contacts. Because there is some confusion in the literature concerning the significance of the Carnot contribution with respect to the second law of Thermodynamics, let us quote Zemansky and Dittman:\cite{PRE9} ``Carnot used \emph{chaleur} when referring to heat in general, but when referring to the motive power of fire that is brought about when heat enters an engine at high temperature and leaves at low temperature, he uses the expression \emph{chute de calorique}, never \emph{chute de chaleur} [\ldots].  Carnot had in the back of his mind the concept of entropy, for which he reserved the term \emph{calorique}.''

\begin{acknowledgments}
The authors thank the anonymous referees for sharp reading and useful comments. 
\end{acknowledgments}


\begin{thebibliography}{9}

\bibitem{Ehrenfest1907}
P.~Ehrenfest and T.~Ehrenfest,
 ``Ueber zwei bekannte {E}ingew{\"a}nde gegen das {B}oltzmannsche
  {H}-{T}heorem,'' 
  Zeitschrift f{\"u}r Physik {\bf 8}, 311--314 (1907).

\bibitem{Guemez1989}
J.~G{\"u}{\'e}mez, S.~Velasco, and A.~Calvo~Hern{\'a}ndez, 
 ``A generalization of the {E}hrenfest urn model,'' 
 Am. J. Phys. {\bf 57}, 828--834 (1989).

\bibitem{ISI:000253989000010}
J.~Tobochnik and H.~Gould, 
 ``Teaching statistical physics by thinking about models and algorithms,'' 
 Am. J. Phys. {\bf 76}, 353--359 (2008).

\bibitem{Quan2009}
H.~T. Quan,
 ``Quantum thermodynamic cycles and quantum heat engines {(II)},'' 
Phys. Rev. E {\bf 79}, 041129 (2009).

\bibitem{PRE9}
M.~Zemansky and R.~Dittman, 
\textsl{Heat and Thermodynamics} 
 (MacGraw-Hill, New-York, 1997).

\bibitem{ourPRE}
J.~Arnaud, L.~Chusseau, and F.~Philippe, 
 ``Mechanical equivalent of quantum heat engines,'' 
Phys. Rev. E {\bf 77}, 061102 (2008).

\bibitem{ejp02}
J.~Arnaud, L.~Chusseau, and F.~Philippe, 
 ``Carnot cycle for an oscillator,''
 Eur. J. Phys. {\bf 23}, 489--500 (2002).

\bibitem{Kastler1976}
A.~Kastler, 
\textsl{Sadi Carnot et l'essor de la Thermodynamique}  (Editions du CNRS, Paris, 1976), 
chapter ``L'\oe uvre posthume de Sadi Carnot,'' page 195.

\bibitem{PRE10}
V.~M. Brodiansky,
\textsl{Sadi Carnot}
 (Presses Universitaires de Perpignan, Perpignan, 2006), 
 translated from Russian to French.

\bibitem{Mer1954}
V.~K. La~Mer,
 ``Some current misinterpretations of {N}. {L}. {S}adi {C}arnot's memoir
  and cycle,''
 Am. J. Phys. {\bf 22}, 20--26 (1954).

\bibitem{hoyer}
U.~Hoyer, 
 ``How did {C}arnot calculate the mechanical equivalent of heat?,''
 Centaurus {\bf 19}(3), 207--219 (1975).

\bibitem{EJP3}
N.~S. Carnot, 
\textsl{R{\'e}flexions sur la puissance motrice du feu}
 (Bachelier, Paris, 1824). 
 Facsimile of the original edition by Jacques Gabay ed., Sceaux, 1960. English translation by Dover publ. Inc., New-York, 1960.

\end{thebibliography}
\end{document}